\documentclass[a4paper,10pt]{article}
\usepackage{amsmath,amscd,amssymb}
\usepackage{epsfig}

\newcommand{\ID}{\mbox{{\sf 1}\hspace{-0.55mm}\rule{0.04em}{1.53ex}}}

\newcommand{\gapeq}{\stackrel{\scriptscriptstyle\raisebox{-2.5mm}{$>$}}
{\scriptscriptstyle \raisebox{-0.8mm}{$\sim$}}}

\begin{document}
\title{\bf\huge Magnetic Moments of Baryons with a Single Heavy Quark}
\author{Stephan Scholl\\Institut f\"ur theoretische Physik,
Eberhard-Karls Universit\"at T\"ubingen\\Auf der Morgenstelle 14,
D-72076 T\"ubingen, Germany\and Herbert Weigel\\Fachbereich Physik,
Universit\"at Siegen\\Walter Flex Stra{\ss}e 3, 57072 Siegen}

\maketitle
\begin{abstract}We calculate the magnetic moments of heavy baryons 
with a single heavy quark in the bound-state approach. In this approach 
the heavy baryons is considered as a heavy meson bound in the 
field of a light baryon. The light baryon field is represented as a 
soliton excitation of the light pseudoscalar and vector meson fields. 
For these calculations we adopt a model that is both chirally invariant 
and consistent with the heavy quark spin symmetry. We gauge the model
action with respect to photon field in 
order to extract the electromagnetic current operator and obtain
the magnetic moments by computing pertinent matrix elements of this
operator between the bound state wavefunctions. We compare our 
predictions for the magnetic moments with results of alternative 
approaches for the description of heavy baryon properties.
\end{abstract}

\section{Introduction}

There has been recent interest to study properties of baryons
that contain heavy quarks. In particular the development of the 
heavy quark or Isgur-Wise symmetry~\cite{Ei81} has generated many 
studies~\cite{Je92}\nocite{Rho93,Gu93,Mo94,Oh94}--\cite{Oh94a} on 
the spectrum of baryons with a single heavy quark in the so--called
bound state approach. To set the notation for what we consider
a baryon with a single heavy quark we denote heavy 
quarks ({\it charm} or {\it bottom}) by ``$Q$'' and light quarks
({\it up} or {\it down}) by ``$q$''. In the quark language a 
baryon whose properties we wish to explore has the structure $qqQ$. 
This defines the quantum numbers of the considered baryon. 
In general, of course, the structure of such a baryon may be more
complicated because additional quark--antiquark--pairs may be excited.
In the above mentioned bound state approach the
heavy baryon is considered as a heavy spin multiplet of mesons (with
structure $Q\bar{q}$) bound in the field of the nucleon, or more
generally a light baryon of $qqq$--structure.
As  suggested by the $1/N_{\rm C}$ expansion~\cite{tHo74,Wi79} of QCD
the field of the nucleon emerges as a soliton configuration of light 
meson fields of $q\bar{q}$ structure. Reviews of the soliton approach to 
the light baryons are given in 
refs.~\cite{Ho86}\nocite{Me88,Sch89,Al96}--\cite{We96}
while the bound state approach for the ``light'' hyperons is discussed
in refs.~\cite{Ca85}\nocite{Ku89,Oh91}--\cite{Bl88}.

The basic idea is to consider an effective meson model that includes 
both, light degrees of freedom and mesons that contain a heavy quark. 
In such a model the light components of the heavy meson fields
couple the light meson fields according to the rules of chiral 
symmetry. Thus this it permits not only an expansion of heavy baryon 
properties in powers of $1/M$, $1/N_{\rm C}$~\cite{Az01} but also in the 
number of derivatives acting on the light components of the heavy system.
However, this introduces a large number of unknown parameters in the
model Lagrangian. We therefore find it more compelling to employ
a model with light vector mesons ($\rho$ and $\omega$) to construct
the soliton rather than a model with many derivatives of the light 
pseudoscalar fields. Actually the need for light vector mesons is 
not surprising since, in the soliton approach, they are necessary to 
explain, for example, the neutron--proton mass difference~\cite{Ja89},
the nucleon axial singlet matrix element~\cite{Jo90} and the 
``high--energy'' behavior of phase shifts in meson--baryon 
scattering~\cite{Sch89}. Furthermore heavy quark components of the 
heavy meson field are required to exhibit the heavy quark symmetry as 
the masses tend to infinity, {\it i.e.} spin and flavor independent 
interactions. Therefore the model also requires both heavy pseudoscalar 
and heavy vector meson fields. Based on a model~\cite{Sch93} that
reflects all those features, the heavy baryon mass splittings have been 
discussed~\cite{Sch95a}, obtaining satisfactory agreement with experiment. 
Also pentaquarks of structure $qqq{\bar Q}q$ have been considered in
this model as antimesons (${\bar Q}q$) bound in the field of the soliton.

Here we will go one step further and study magnetic moments of baryons 
with a single heavy quark in the model suggested in ref.~\cite{Sch93} as 
an example for computing static properties. This not only requires
to construct the soliton of the light meson fields, to compute the heavy 
meson bound state wavefunction, and to quantize the the combined system 
to generate states to be identified as heavy baryons but in addition also 
to derive of the electromagnetic current 19.12. operator
in the model and to subsequently evaluate matrix elements thereof between
heavy baryon states. It is important to note that both the light
and the heavy sector of the model will contribute to that current 
and hence to the magnetic moments of heavy baryons. Stated otherwise,
these magnetic moments directly vary with the parameters of the light 
sector and not only indirectly via the soliton profiles as {\it e.g.}
the spectrum of the heavy baryons does.

This paper is organized as follows. In section 2 we will review
the model Lagrangian for both the light and heavy flavor sectors
of the model. This will also provide the opportunity to derive
covariant expressions for the electromagnetic current.
Section 2 also contains a brief discussion the soliton profiles
that emerge in the light flavor sector as well the heavy meson
bound state profiles. In section 3 we will generate states with
good baryon quantum numbers by canonically quantizing collective
coordinates that parameterize the (zero--mode) rotations of 
the combined soliton -- heavy meson system. This makes mandatory
the discussion of field components that are absent classically
but get induced by the collective rotation. Section 4 represents
the major progress for the studies of heavy baryons as a heavy
meson bound to a soliton as we construct the magnetic moment
operator in this model and compute the corresponding matrix elements
numerically. Concluding remarks are to be found in section~5. Also
we complete the paper by including an appendix that summarizes
functionals of the meson profiles that emerge along the computation.

Some preliminary results of this study have already been 
presented in a conference contribution~\cite{We03}.

\section{The model Lagrangian}
The model Lagrangian can be divided in two distinct parts. The first 
part, ${\cal L}_{\rm light}$ describes the interaction of the mesons that
are built out of the light quarks {\it up} and {\it down}. 
Our model not 
only contains the pseudoscalar pion fields but also the vector mesons 
$\omega$ and $\rho$. The interactions among these light mesons is dictated 
by chiral symmetry. The second part, ${\cal L}_{\rm heavy}$
in addition contains fields that correspond to mesons that
are built out of a single heavy quark ({\it charm} or {\it bottom})
and light quarks\footnote{In what follows we will
refer to mesons with a single heavy quark as heavy mesons.}.
This part of the Lagrangian is consistent with the heavy quark spin 
flavor symmetry and therefore contains both pseudoscalar
and vector degrees of freedom. In the charm quark sector these are
$D(1865)$ and $D^\ast(2007)$ while $B(5279)$ and $B^\ast(5325)$ contain a 
bottom quark.

The Lagrangian ${\cal L}_{\rm light}$ contains (topological) 
soliton solutions that are identified with baryons according to
the Skyrme--model picture. Substituting this soliton configuration
for the light meson fields in ${\cal L}_{\rm heavy}$ generates a potential
for the heavy meson fields. The Klein--Gordon equation associated
with this potential has solutions with energy less than the rest 
mass of the considered heavy mesons, {\it i.e.} a bound heavy meson.
We combine bound heavy mesons with the soliton of the light 
mesons ({\it i.e.} the soliton represents a baryon of light quarks)
to obtain representatives for baryons with a single heavy quark.
In the soliton description this model for the heavy baryon is that
of a heavy meson bound to a light baryon. In this section we will
review how this picture emerges and also discuss the basics for
computing the magnetic moments of such heavy baryons. 

\subsection[The light meson Lagrangian]{The light meson Lagrangian}

The light sector of the model is given by the chirally invariant Lagrangian 
discussed in refs.~\cite{Ja88,Ka84}. 
In addition to the standard Skyrme
model, which considers pseudoscalar pion fields only, this Lagrangian
also contains vector meson degrees of freedom. In order to efficiently
incorporate chiral symmetry it is most useful to employ the 
non--linear representation of the isovector pion fields ($\vec{\pi}$)
via the chiral field 
$U={\rm exp}\left(i\vec{\pi}\cdot\vec{\tau}/f_\pi\right)$ where 
$f_\pi=93{\rm MeV}$ and $\vec{\tau}$ are the pion decay constant
and the Pauli matrices, respectively. To begin with, a chirally
invariant Lagrangian for vector mesons should contain both,
vector and axialvector fields. The chirally covariant elimination of 
the axialvector introduces the root, $\xi$ of the chiral field, {\it i.e.}
$\xi^2=U$~\cite{Ka84} as well as vector and axialvector currents
of the pion fields:
\begin{equation}
v_{\mu} = \frac{i}{2}
\left(\xi{\partial_{\mu}}{\xi^{\dagger}}
+{\xi^{\dagger}}{\partial_{\mu}}{\xi}\right),\qquad
p_{\mu}   = \frac{i}{2}
\left(\xi{\partial_{\mu}}{\xi^{\dagger}}
-{\xi^{\dagger}}{\partial_{\mu}}{\xi}\right)\,.
\label{defpv}
\end{equation}
We comprise the isoscalar $\omega$ and the isovector $\vec{\rho}$
within a single matrix field
$\rho_{\mu}=(\vec{\rho\,}_\mu\cdot\vec{\tau}+\omega_{\mu}{\ID})/2$
and further define
\begin{equation}
F_{\mu\nu}(\rho) =
\partial_{\mu}\rho_{\nu}-\partial_{\nu}\rho_{\mu}
+ig[\rho_\mu,\rho_{\nu}]\qquad {\rm and}\qquad
R_{\mu}  =  \rho_{\mu}-\frac{1}{g}v_{\mu}
\label{defFR}
\end{equation}
to compactly write the normal parity part of the Lagrangian 
for the light mesons
\begin{equation}
{\cal L}_S=f_\pi^2{\rm tr}\left[p_{\mu}p^{\mu}\right]
+\frac{1}{4}m_\pi^2f_\pi^2{\rm tr}\left[U+U^{\dagger}-2\right]
-\frac{1}{2}{\rm tr}\left[F_{\mu\nu}(\rho)F^{\mu\nu}(\rho)\right]
+m_{\rm V}^2{\rm tr}\left[R_{\mu}R^{\mu}\right]\,.
\label{lRPLag}
\end{equation}
Here $m_\pi=138{\rm MeV}$ and $m_{\rm V}=770{\rm MeV}$ are pion and 
vector meson masses, respectively. The last term in eq.~(\ref{lRPLag})
contains the tri--linear vertex 
$\vec{\rho}_\mu\cdot\left(\vec{\pi}\times\partial^\mu\vec{\pi}\right)$
and allows to determine the coupling constant $g\approx5.6$ from the
decay $\rho\to \pi\pi$. In addition the light meson Lagrangian
contains a part that involves the Levi-Cevita tenor
$\epsilon_{\mu\nu\alpha\beta}$. Since this part includes the nonlocal
Wess--Zumino--Witten term it is most convenient to write it in 
action form using  differential forms like $p=p_\mu dx^\mu$
\begin{eqnarray}
\Gamma_{an}(p,R,F)&=&\frac{2N_C}{15\pi^2}\int
{\rm tr}[p^{5}]\nonumber\\
&&+ \int
{\rm tr}\left[\frac{4i}{3}\left(\gamma_1+\frac{3}{2}\gamma_2\right)Rp^3
-\frac{g}{2}\gamma_2F(\rho)(pR-Rp)-2ig^2(\gamma_2+2\gamma_3)
R^{3}p\right].\nonumber\label{lanA}
\end{eqnarray}
Note that we will only consider the physical case of $N_C=3$ color
degrees of freedom.

In ref.~\cite{Ja88} two of the three unknown constants $\gamma_{1,2,3}$ were
determined from purely strong interaction processes. Defining
$\tilde h=-\frac{2\sqrt2}{3}\gamma_1$, $\tilde g_{VV\phi}=g\gamma_2$ and
$\kappa=\frac{\gamma_3}{\gamma_2}$ the central values $\tilde h=0.4$ and
$\tilde g_{VV\phi}=1.9$ were found. Within experimental uncertainties
(stemming from the uncertainty in the $\omega - \phi$ mixing angle) these
may vary in the range $\tilde h=-0.15,..,0.7$ and $\tilde g_{VV\phi}=1.3,..,
2.2$ subject to the constraint $\vert\tilde g_{VV\phi}-\tilde h\vert\approx
1.5$. The third parameter, $\kappa$ could not be fixed in the meson sector,
however, from the study \cite{Ja88} of nucleon properties in the $U(2)$ reduction 
of the model it was argued that $\kappa\approx1$. To be specific, we will
always employ the parameters:
\begin{equation}
g=5.6\,,\quad \tilde{h}=0.3\,,\quad
\tilde{g}_{{\rm VV}\phi}=1.8\,,\quad \kappa=1.2\,.
\label{parameters}
\end{equation}
that have been found suitable for describing the low--lying baryons
for three light flavors \cite{Pa92}.

The total action for the light mesons is the sum
\begin{equation}
\Gamma_{\rm light}[\xi,\rho]=\int d^4x {\cal L}_S +\Gamma_{\rm an}\,.
\label{lightaction}
\end{equation}

The electromagnetic current associated with $\Gamma_{\rm light}$ 
is most easily obtained by gauging it with the photon field
${\cal A}_\mu\,{\cal Q}$ where ${\cal Q}={\rm diag}(2/3,-1/3)$
is the quark charge matrix and extracting the linear term:
\begin{equation}
{\cal J}_\mu^{\rm light}=\frac{\delta\Gamma[\xi,\rho,{\cal A}]}
{\delta{\cal A}^\mu}\Big|_{{\cal A}=0}\,.
\label{currentextraction}
\end{equation}
This yields
\begin{eqnarray}
{\cal J}_\mu^{\rm light}&=&
f_\pi^2{\rm tr}\left\{{\cal Q}\left[
\xi^\dagger p_\mu\xi-\xi p_\mu\xi^\dagger\right]\right\}
-\frac{m_V^2}{g}{\rm tr}\left\{{\cal Q}\left[
\xi R_\mu\xi^\dagger+\xi^\dagger R_\mu\xi\right]\right\}\cr
&&-\frac{2i}{3}\left[\frac{N_c}{4\pi^2}-\frac{1}{g}
\left(\gamma_1+\frac{3}{2}\gamma_2\right)\right]
\epsilon_{\mu\nu\rho\sigma}{\rm tr}\left\{\left[
\xi^\dagger {\cal Q}\xi+\xi {\cal Q}\xi^\dagger\right]
p^\nu p^\rho p^\sigma\right\}\cr
&&-\frac{2i}{3}\left(\gamma_1+\frac{3}{2}\gamma_2\right)
\epsilon_{\mu\nu\rho\sigma}
{\rm tr}\left\{\left[\xi^\dagger{\cal Q}\xi-\xi{\cal Q}\xi^\dagger\right]
\left[R^\nu p^\rho p^\sigma+p^\nu p^\rho R^\sigma
-p^\nu R^\rho p^\sigma\right]\right\}\cr
&&-\frac{g}{4}\gamma_2\epsilon_{\mu\nu\rho\sigma}
{\rm tr}\left\{{\cal Q}\left(
\xi\left[F^{\nu\rho}R^\sigma+R^\nu F^{\rho\sigma}
-\frac{1}{g}\left(F^{\nu\rho}p^\sigma+p^\nu 
F^{\rho\sigma}\right)\right]\xi^\dagger
\right.\right.
\cr &&
\left.\left.
\qquad\qquad\mp\xi^\dagger\left[F^{\nu\rho}R^\sigma+R^\nu F^{\rho\sigma}
+\frac{1}{g}\left(F^{\nu\sigma}p^\rho+p^\nu F^{\rho\sigma}\right)\right]
\xi\right)\right\}\cr
&&-ig^2\left(\gamma_2+2\gamma_3\right)\epsilon_{\mu\nu\rho\sigma}
{\rm tr}\left\{{\cal Q}
\left(\xi\left[R^\nu R^\rho R^\sigma
-\frac{1}{g}\left(R^\nu R^\rho p^\sigma+
p^\nu R^\rho R^\sigma-R^\nu p^\rho R^\sigma\right)\right]\xi^\dagger
\right.\right.\cr &&\left.\left.
\qquad\qquad-\xi^\dagger\left[R^\nu R^\rho R^\sigma
+\frac{1}{g}(R^\nu R^\rho p^\sigma+p^\nu R^\rho R^\sigma
-R^\nu p^\rho R^\sigma)\right]\xi\right)\right\}\cr
&&+2d_1\epsilon_{\mu\nu\rho\sigma}\partial^\nu\{
{\rm tr}\left\{{\cal Q}\left[\xi\left(
R^\rho p^\sigma-p^\rho R^\sigma\right)\xi^\dagger
+\xi^\dagger\left(R^\rho p^\sigma
-p^\rho R^\sigma\right)\xi\right]\right\}\,.
\label{lightcurrent}
\end{eqnarray}
The last term, proportional to $d_1$ has no analogue in the pure hadronic 
part of the action~(\ref{lanA}). This term is of electromagnetic
nature and the coupling constant can be related to the light vector 
meson decay $\omega\rightarrow\pi^{0}\gamma$:
$|2d_1-\frac{\gamma_2}{2}|=0.038\pm0.002$~\cite{Me89}.

\subsection{Solitons}

The action for the light degrees of freedom contains static solition
solutions. The hedgehog type configuration reads~\cite{Ja88}:
\begin{eqnarray}
U(\vec{r\,})&=& \exp\left(i\vec{r}\cdot\vec{\tau\,}F(r)\right)\, 
\quad , {\it i.e.}\quad
\xi(\vec{r\,})=\exp\left(i\vec{r}\cdot\vec{\tau\,}\frac{F(r)}{2}\right)\,\cr
\rho^a_{i}(\vec{r\,})&=&\frac{G(r)}{gr}
\epsilon_{ija}\hat{r}_j\tau_a\,,
\quad {\rm and}\quad
\omega(\vec{r\,}) = \frac{\omega(r)}{g}\,,
\label{Sol} 
\end{eqnarray}
while all other field components vanish classically. 
These configurations are invariant under combined spatial and
flavor rotations generated by the vector sum $\vec{L}+\vec{S}+\vec{I}$,
the so--called grand spin. Assuming the boundary conditions
(a prime indicates the derivative with respect to $r$)
\begin{equation}
F(0)=\pi\,,\quad G(0)=-2\,,\quad \omega^\prime(0)=0\,,\quad
F(\infty)=G(\infty)=\omega(\infty)=0
\label{bc}
\end{equation}
yields baryon number one solitons by substituting the 
{\it ans\"atze}~(\ref{Sol}) into the action~(\ref{lightaction})
and extremizing the resulting functional of the classical mass, 
$M_{\rm cl}[F,G,\omega]$. For completeness we list the explicit form
of that functional in the appendix. Note, that this soliton
does not yet describe baryon states with good spin and flavor quantum
numbers. Before describing the cranking procedure that generates
such states, we will focus on that part of the model Lagrangian
that contains the heavy meson fields.
  
\subsection[The relativistic Lagrangian for the heavy mesons]{The
relativistic Lagrangian for the heavy mesons}

The relativistic Lagrangian ${\cal L}_{\rm heavy}$ which describes the
coupling between light and heavy mesons is given by \cite{Sch93}:
\begin{eqnarray}
{\cal L}_{\rm heavy} &  = & D_\mu P(D^\mu P)^\dagger
-\frac{1}{2}Q_{\mu\nu}(Q^{\mu\nu})^\dagger
-M^{2}PP^\dagger+M^{\ast2}Q_\mu Q^{\mu\dagger}\cr
&+&2iMd\left(Pp_\mu Q^{\mu\dagger}-Q_\mu p^\mu P^\dagger\right)
-\frac{d}{2}\epsilon^{\alpha\beta\mu\nu}
\left[Q_{\nu\alpha}p_\mu Q_\beta^\dagger
+Q_\beta p_\mu (Q_{\nu\alpha})^\dagger\right]\cr
&-&\frac{2\sqrt{2}icM}{m_{V}}
\left\{2Q_\mu F^{\mu\nu}(\rho)Q_\nu^\dagger
-\frac{i}{M}\epsilon^{\alpha\beta\mu\nu}
\left[D_\beta PF_{\mu\nu}(\rho)Q_\alpha^\dagger
+Q_\alpha F_{\mu\nu}(\rho)(D_\beta P)^\dagger\right]\right\}\,.
\label{hLag}
\end{eqnarray}
Here the mass $M$ of the heavy pseudoscalar Meson $P$ differs from the
mass $M^{\ast}$ of the heavy vector meson $Q_{\mu}$. We take
\begin{eqnarray}
M&=& 1865\: MeV\, {\rm (D-meson)}\,,
\quad 5279\: MeV\, {\rm (B-meson)},\cr
M^{\ast}&=& 2007\: MeV\, {\rm (D-meson)}\,,
\quad 5325\: MeV\, {\rm (B-meson)}\,.
\label{hPar1}
\end{eqnarray}

It should be
noticed that the heavy meson fields are conventionally defined as row
vectors in isospin space. The chirally covariant derivatives of the
heavy fields $P$ and $Q$ and the field-strength tensor $Q_{\mu\nu}$
are defined as:
\begin{eqnarray}
D_\mu P^\dagger&=&\left[\partial_\mu-i\alpha g\rho_{\mu}
-i\left(1-\alpha\right)v_\mu\right]P^\dagger
=\left[\partial_\mu-iv_\mu-ig\alpha R_\mu\right]P^\dagger\,,\cr\cr
D_\mu Q_\nu^\dagger&=&\left[\partial_\mu-iv_\mu
-ig\alpha R_\mu\right]Q_\nu^\dagger\,,\quad
(Q_{\mu\nu})^\dagger=D_\mu Q_\nu^\dagger-D_\nu Q_\mu^\dagger.
\label{hchicovarD}
\end{eqnarray}
In eq.~(\ref{hLag}) we have two sets of two terms each that involve
the identical coupling constants $d$ and~$c$. Each of these terms
is chirally invariant and could thus in principle carry independent
coupling constants. However, the condition that the Lagrangian 
${\cal L}_{\rm heavy}$ obeys the heavy quark spin symmetry in the
limit $M,M^\ast\to\infty$ requires the combination of these terms
as given in  eq.~(\ref{hLag})~\cite{Sch93}. Note however, that we 
do not assume infinitely large masses for the heavy meson. 
The coupling constants $d$, $c$ and $\alpha$ are still not
precisely determined. While $d$ and $c$ can be determined from the
decay widths of the heavy mesons~\cite{Ja95}
\begin{equation}
d= 0.53\,,\qquad c = 1.60 \,,
\label{hPar2}
\end{equation}
there is no
direct experimental information for the value of $\alpha$. The value of
$\alpha$ would be unity if a possible definition of light vector meson
dominance for the electromagnetic form factors of the heavy meson was
adopted \cite{Ja95}.

Due to the electromagnetic gauging of the Lagrangian extra terms appear 
in the light vector fields $\rho_\mu$, $v_\mu$ and $p_\mu$. The respective
electromagnetic and chirally invariant forms read (we choose the
electric charge to be unity):
\begin{equation}
\tilde{\rho}_\mu=\rho_\mu-\frac{1}{g}{\cal Q}{\cal A}_\mu\,,\quad
\tilde{v}_\mu=v_\mu+\frac{1}{2}{\cal A}_\mu
\left(\xi{\cal Q}\xi^{\dagger}+\xi^{\dagger}{\cal Q}\xi
-2{\cal Q}\right)\quad{\rm and}\quad
\tilde{p}_\mu=p_\mu-\frac{1}{2}{\cal A}_\mu
\left(\xi{\cal Q}\xi^{\dagger}-\xi^{\dagger}{\cal Q}\xi\right)\,.
\label{echirho}
\end{equation}
Considering these extra terms, the electromagnetic and chiral
covariant derivative of the heavy meson fields has to be modified:
\begin{eqnarray}
D_{\mu}P^\dagger&=&\left(\partial_{\mu}-i\alpha g\tilde{\rho}_\mu
-i\left(1-\alpha\right)\tilde{v}_\mu-i\left({\cal Q}-{\cal C}\right)
{\cal A}_{\mu}\right)P^\dagger\cr
&=&\left(\partial_{\mu}-i\alpha
g\rho_\mu-i\left(1-\alpha\right)v_\mu+ie{\cal A}_\mu\left[{\cal
C}-\frac{1}{2}\left(1-\alpha\right)
\left(\xi{\cal Q}\xi^{\dagger}+\xi^{\dagger}{\cal
Q}\xi\right)\right]\right)P^\dagger\,,
\label{hechiD}
\end{eqnarray}
which makes plausible the above assertion that $\alpha=1$ 
corresponds to vector meson dominance because for that value there
is only the direct coupling to the heavy mesons.
The charge of the heavy quark in question is denoted by $\cal C$,
{\it i.e.} ${\cal C} = \frac{2}{3}$ for the charm sector and ${\cal C} =
-\frac{1}{3}$ for the bottom sector. Finally we substitute 
eq.~(\ref{hechiD}), for $D_{\mu}P^\dagger$, its analogue
for $(D_\mu Q_\mu)^\dagger$ as well as $\tilde{\rho}$,
etc. from eq.~(\ref{echirho}) for $\rho$ into the Lagrangian~(\ref{hLag})
to obtain the electromagnetic current of the heavy mesons
\begin{eqnarray}
{\cal J}_\mu^{\rm heavy}=\left. 
\frac{\delta {\cal L}_{\rm heavy}}{\delta{\cal A}^{\mu}}\right
|_{{\cal A}_{\mu}=0}&=&
i\left(P\tilde{C}\left(D_{\mu}P\right)^\dagger
-D_\mu P\tilde{C}P^\dagger\right)
+i\left(Q^{\nu}\tilde{C}Q_{\mu\nu}^{\dagger}
-Q_{\mu\nu}\tilde{C}Q^{\nu\dagger}\right)\cr
&&-ieMd\left(P\left(\xi{\cal Q}\xi^\dagger
-\xi^\dagger{\cal Q}\xi\right)Q_\mu^\dagger-Q_\mu
\left(\xi{\cal Q}\xi^\dagger-\xi^\dagger{\cal Q}\xi\right)P\right)\cr
&&-id\epsilon_{\alpha\beta\mu\nu}
\left(Q^\alpha\tilde{C}p^\nu Q^{\beta\dagger}
-Q^\beta p^\nu\tilde{C}Q^{\alpha\dagger}\right.\cr
&&\quad\left.+ \frac{i}{4}\left[Q^{\nu\alpha}
\left(\xi{\cal Q}\xi^{\dagger}-\xi^{\dagger}{\cal Q}\xi\right)
Q^{\beta\dagger}+Q^\beta\left(\xi{\cal Q}\xi^\dagger
-\xi^\dagger{\cal Q}\xi\right)Q^{\nu\alpha\dagger}\right]\right)\cr
&&+i\frac{2\sqrt{2}c}{m_V}\epsilon_{\alpha\beta\mu\nu}
\left(P\tilde{C}F^{\beta\nu}(\rho)Q^{\alpha\dagger}
-Q^\alpha F^{\beta\nu}(\rho)\tilde{C}P^{\dagger}\right)\,.
\label{Jh}
\end{eqnarray}
Here the abbreviation $\tilde{C} = {\cal C}- \frac{1-\alpha}{2}\left(\xi{\cal
Q}\xi^{\dagger}+\xi^{\dagger}{\cal Q}\xi\right)$ has been employed.
A similar electromagnetic current was also derived in ref.~\cite{Ja95}
to discuss radiative decays of heavy vector mesons. However, the $d$--type
term in eq.~(\ref{Jh}) was omitted as it does not contribute to
such processes.

\subsection[Bound states]{Bound states}

Substituting the soliton configuration~(\ref{Sol}) for the light 
meson fields in ${\cal L}_{\rm heavy}$ generates a potential for
the heavy meson fields. The existence of heavy meson bound states 
in this potential has been established some time ago \cite{Sch95a}. 
Such bound states emerge for both, positive and negative frequency
modes. The former correspond to mesons bound to the light quark
soliton and thus carry quantum numbers of ordinary three--quark baryons.
The latter, however, have antimesons bound to the soliton and
their quantum numbers cannot be constructed in a three--quark 
picture. Rather additional quark--antiquark excitations are required
and the resulting heavy baryons possess pentaquark structures.

In the current study we want to compute the 
magnetic moments of ground state baryons with a single heavy
quark, {\it i.e.} $\Lambda_c, \Sigma_c$ when the heavy meson is 
a D--meson and $\Lambda_b, \Sigma_b$ in the case of the B--meson. 
The ground states, of course, correspond to the most tightly bound
state. Due to the hedgehog structure of the background soliton,
which has nonzero orbital angular momentum, these bound states
emerge in the $P$--wave channels\footnote{The orbital angular momentum 
quantum numbers denote those of the pseudoscalar component 
$P^{\dagger}$ of the heavy meson multiplet 
$(P^{\dagger},Q_{\mu}^{\dagger})$.}, rather than in the $S$--wave
as na{\"\i}vly assumed. For that reason we will only consider
the $P$--wave channel and refer to ref.~\cite{Sch95a} for details
on $S$--wave channel bound states. The appropriate {\it ansatz}
for the heavy meson multiplet in the P-wave channel reads:
\begin{eqnarray}
P^\dagger(\vec{r\,},t) & = &\frac{1}{\sqrt{4\pi}}\Phi(r)
\hat{r}\cdot\vec{\tau}\,\rho(\epsilon )\,{\rm e}^{i\epsilon t}\,,\quad
Q_0^\dagger(\vec{r\,},t)=\frac{1}{\sqrt{4\pi}}\Psi_0(r)\,\rho(\epsilon )\,
{\rm e}^{i\epsilon t}\,,\cr
Q_i^\dagger(\vec{r\,},t)&=&\frac{1}{\sqrt{4\pi}}
\left[i\Psi_1(r)\hat{r}_i+\frac{1}{2}\Psi_2(r)
\epsilon_{ijk}\hat{r}_j\tau_k\right]\,\rho(\epsilon )\,
{\rm e}^{i\epsilon t}\,.
\label{PQi}  
\end{eqnarray}
Here $\rho(\epsilon)=\begin{pmatrix}
\rho_1(\epsilon)\cr\rho_2(\epsilon)\end{pmatrix}$ 
refers to a properly normalized 
isospinor describing the isospin of the the heavy meson multiplet. 
Upon canonical quantization the Fourier
amplitudes $\rho(\epsilon)$ and $\rho^{\dagger}(\epsilon)$ are
elevated to annihilation and creation operators for a heavy
meson quantum with the energy eigenvalue $\epsilon$ (see below).
We substitute the field configurations (\ref{Sol}) and (\ref{PQi}) 
into the Lagrangian given by eqs.~(\ref{lightaction}) and (\ref{hLag})
and integrate over coordinates to obtain the Lagrange function
of the form
\begin{equation}
L = -M_{cl}[F,G,\omega]+
I_{\epsilon}[F,G,\omega;\Phi,\Psi_{0},\Psi_{1},
\Psi_{2}]\rho^{\dagger}(\epsilon )\rho(\epsilon)\,.
\label{BSLag}
\end{equation}
Here $M_{cl}$ denotes the classical solition mass. Its variation yields 
the profile functions $F$, $G$ and $\omega$. The heavy meson
fields are contained in the functional $I_{\epsilon}$ where the 
subscript indicates the parameterical dependence on the frequency 
of the fluctuation heavy meson fields. The explicit expressions for 
the functionals $M_{cl}$ and $I_{\epsilon}$ are given in the appendix.
We vary $I_{\epsilon}$ with respect to $\Phi,\Psi_{0},\Psi_{1}$
and $\Psi_{2}$ to get the equations of motions for the heavy
meson fields. These constitute Klein--Gordon type equations with 
potentials generated by the profile functions $F$, $G$ and $\omega$.
We then tune the frequency $\epsilon$ such that these equations
yield a normalizable solution with $|\epsilon| < M$. The solution
with the smallest such $|\epsilon|$ is the bound state we are 
looking for. The equations of motions for the heavy meson fields
are homogeneous linear differential equations. Hence they
provide a solution only up to an overall prefactor. Nonetheless, the 
equations of motion for the heavy meson
fields allow us to extract a metric for a scalar product between
different bound states. In particular its diagonal elements can be
used to properly normalize the bound state wave functions. The 
physical interpretation of the normalization condition is that each
occupation of the bound state should add the amount $|\epsilon|$ to
the total energy and that such a bound states carries unit heavy
flavor. Thus we obtain the normalization condition
\begin{equation}
\left|\frac{\partial}{\partial{\epsilon}}
I_{\epsilon}[\Phi,\Psi_{0},\Psi_{1},\Psi_{2}]\right|= 1\label{PNorm}
\end{equation}
in addition to the canonical commutation relation
$[\rho_i(\epsilon),\rho_j^{\dagger}(\epsilon^{\prime})]
=\delta_{ij}\delta_{\epsilon,\epsilon^{\prime}}$ for the 
Fourier amplitude of the bound state. For further details we
again refer to ref.~\cite{Sch95a}.

\section{States with Spin and Flavor Quantum Numbers}

It is easy to verify that the field configurations for both the
light mesons and the heavy mesons are neither eigenfunctions of the
spin nor the isospin generators. It is the aim of this section to
construct such eigenstates for the bound state system.

\subsection[Collective coordinates]{Collective coordinates}

We employ the {\it cranking} procedure in order to generate states 
that correspond to physical baryons. In a first step we introduce  
collective coordinates that parameterize the (iso-) spin orientation of the meson
configuration,
\begin{equation}
\xi(\vec{r\,},t) = A(t)\xi_{H}(\vec{r\,})A^{\dagger}(t),\qquad 
\rho_{i}(\vec{r\,},t) = A(t)\rho_{i\:H}(\vec{r\,})A^{\dagger}(t).\label{rhoR}
\end{equation}
The time dependence of the collective rotations is measured by the
angular velocities $\vec{\Omega\,}$:
\begin{equation}
A^{\dagger}(t)\frac{d}{dt}A(t)=\frac{i}{2}
\vec{\tau}\cdot\vec{\Omega}\,. 
\label{Omega}
\end{equation}
In addition to the collective rotation of the soliton configuration,
classically vanishing field components are induced. In the case of the
light vector mesons these are given by \cite{Me89}:
\begin{equation}
\omega_{i}=\frac{1}{g}\varphi(r)\epsilon_{ijk}
\Omega_{j}\hat{r}_{k}\,,\quad
\rho_{0}^{k}=\frac{1}{g}\left[\xi_{1}(r)\Omega_{k}
+\xi_{2}(r)\hat{r}\cdot\vec{\Omega}\hat{r}_{k}\right]\,\,.
\label{rhoI}
\end{equation}
After introducing these additional fields the Lagrangian for the light
mesons now contains a term which is quadratic in the angular
velocities. The constant of proportionality defines the moment of
inertia $\alpha^{2}[F,G,\omega;\xi_{1},\xi_{2},\varphi]$.
Varying this moment of inertia with respect to the
radial functions $\xi_{1}(r)$, $\xi_{2}(r)$ and~$\varphi (r)$ 
yields their equations of motion that are linear, inhomogeneous
second order differential equations with the inhomogeneity 
defined by the classical profile functions $F,G$ and~$\omega$.
We solve these differential equations subject to boundary 
conditions that avoid singularities in both the respective 
equations of motion and the moment of inertia~\cite{Me89}.
The moment of inertia as evaluated for this solution is a pure 
number. For the parameters listed above eq~(\ref{parameters}) 
we find $\alpha^{2} = 5.01\, {\rm GeV^{-1}}$.

The heavy meson fields also need to be rotated in isospin space, this
is done analogously to eq.(\ref{rhoR}):
\begin{eqnarray}
P^{\dagger}\rightarrow A(t)P^{\dagger},\qquad 
Q_{\mu}^{\dagger}\rightarrow A(t)Q_{\mu}^{\dagger}\,.
\label{heavyR}
\end{eqnarray}
We have now summarized all the collectively rotating fields
and are in the position to give the Lagrange function for
the collective coordinates
Lagrangian ${\cal L}_{\rm light} + {\cal L}_{\rm heavy}$, 
\begin{equation}
L_{\rm coll}=-M_{cl}+I_\epsilon\rho^\dagger\rho+\frac{\alpha^2}{2}
\vec{\Omega\,}^{2}
+\frac{\chi}{2}\rho^\dagger\vec{\Omega}\cdot\vec{\tau\,}\rho 
\label{Lcoll}
\end{equation}
that is obtained by integrating 
${\cal L}_{\rm light}+{\cal L}_{\rm heavy}$ with the above {\it ans\"atze}
substituted. The new quantity is $\chi$ which describes the coupling 
between the collective rotations and the bound state wave functions. 
For reasons that will become
obvious below, we will refer to $\chi$ as the hyperfine parameter. It is 
a functional of all radial functions, {\it i.e.} it contains both light 
and heavy meson fields and can straightforwardly be computed once
a value for $\alpha$ is chosen, {\it cf.} table~\ref{table_1}.

The quantization of the Lagrangian (\ref{Lcoll}) proceeds along the
bound state approach to the Skyrmion \cite{Ca85}--\nocite{Ku89,Oh91}\cite{Bl88}:
Noether charges for spin and flavor have to be constructed. 
Considering the variation of the fields under infinitesimal symmetry 
transformations, we find for the isospin transformation
transformation: 
\begin{equation}
\left[\Phi,i\frac{\tau_{i}}{2}\right]=
-D_{ij}(A)\frac{\partial\dot{\Phi}}{\partial\Omega_{j}}+...\,.
\label{Isospint} 
\end{equation} 
Here $\Phi$ refers to any of the given iso-rotating
meson fields and the ellipses represent subleading terms in $1/N_{C}$,
e.g. time derivatives of the angular velocities which might arise from
eq. (\ref{rhoI}). Furthermore $D_{ij}(A) = {1/2}\;
tr[\tau_{i}A\tau_{j}A^{\dagger}]$ denotes the adjoint representation
of the collective rotations $A$. It is straightforward to conclude
from eq. (\ref{Isospint}) that the total isospin is related to the
derivative of the Lagrange function with respect to the angular
velocities:
\begin{equation}
I_i=-D_{ij}(A)\frac{\partial L}{\partial\Omega_{j}}\,.
\label{Jsol2I}
\end{equation}
The total spin operator $\vec{J}$ contains the grand spin operator 
$\vec{G}$,
\begin{equation}
G_{i}=J_{i}+D_{ij}^{-1}(A)I_{j}=J_{i}-J_{i}^{sol}\,,
\label{Grandspin}
\end{equation}
with $\vec{J}^{sol} = \partial L/\partial \vec{\Omega}$. As a
consequence of eq. (\ref{Jsol2I}) we obtain the relation:
\begin{equation}
(\vec{J}^{sol})^{\: 2}=\vec{I}^{\,\,2}=I(I+1).
\end{equation}
By construction the light meson fields do not contribute to the grand
spin $\vec{G}$. Also those parts of the heavy meson wave functions (\ref{heavyR}) 
that are placed between the collective coordinates $A$ and the
spinor $\rho$ carry zero grand spin. With the normalization condition
(\ref{PNorm}) one therefore finds for the grandspin operator:
\begin{equation}
\vec{G}= -\rho^{\dagger}\frac{\vec{\tau}}{2}\rho.
\end{equation}
This relation connects the operator multiplying the hyperfine
parameter in the collective Lagrangian in eq. (\ref{Lcoll}) to the spin and
isospin operators. The collective coordinate piece of 
the Hamiltonian is then obtained
from the Legendre transformation:\\
\begin{equation}
H_{\rm coll} = \vec{\Omega}\cdot\vec{J}^{sol}- L_{\rm coll} =
\frac{1}{2\alpha^{2}}\left(\vec{J}^{sol}+\chi\vec{G}\right)^{2}\,.
\label{Hcoll}
\end{equation}
Since the moment of inertia, $\alpha^2$ is of order $N_C$, this operator 
is of order $1/N_C$. Finally the mass formula for baryon $B$ with a single
heavy quark reads:
\begin{equation}
M_B=M_{cl}+|\epsilon|+\frac{1}{2{\alpha^{2}}}
\left[\chi J(J+1)+(1-\chi)I(I+1)\right].
\label{Hmass}
\end{equation}
Contributions of $O(\chi^{2})$ have been omitted for consistency 
since these terms are quartic in the heavy meson wavefunctions and 
have been excluded form the model form the very beginning.
Obviously the parameter $\chi$ characterizes the hyperfine splitting.
In ref.~\cite{Sch95a} it
has been verified that it vanishes in heavy quark limit $M$,
$M^{\ast}\rightarrow \infty$~. This is a consequence of heavy quark 
spin symmetry that predicts hadrons to be degenerate that only differ 
by their spin quantum number.

To constrain the unknown parameter $\alpha$ we compute the mass 
differences of the bound P---wave heavy baryons $\Sigma_{C}$ and 
$\Sigma_{C}^{\ast}$ to $\Lambda_{C}$. In addition, we calculate
the mass differences of $\Lambda_{C}$ to the nucleon and $\Lambda_{B}$.
A more detailed comparison containing also S--Wave bound states
and radially excited states can be found in ref. \cite{Sch95a}.
For quite a range of $\alpha$ fair agreement with the empirical 
data is obtained. The $\Sigma_{C}-\Lambda_{C}$ mass differences
suggest a negative value of $\alpha$ while the experimental mass 
difference between the nucleon and the $\Lambda_{C}$ is reproduced 
for $\alpha \approx 0.4$. The
$\Lambda_{B}-\Lambda_{C}$ mass difference is off by only
about 5\%. 
\begin{table}[htb]
\caption{\label{table_1}Quantities that enter the mass 
formula~(\ref{Hmass}) as functions of the unknown parameter~$\alpha$.
The other model 
parameters are as in eqs.~(\ref{parameters}, \ref{hPar1}) and (\ref{hPar2}).
The binding energy is defined as $\omega=M-|\epsilon|$,
$M$ being the mass of the heavy pseudoscalar meson.
We also compare the resulting mass differences to experimental
data in cases the latter are available~\cite{PDG}.
The superscripts denote the heavy quark flavor under
consideration.  Energies are measured in MeV and masses are
displayed relative to $\Lambda_C$.}
\begin{center}
\begin{tabular}{|r||c|c|c|c|c|c|c|c|c|c||c|} \hline
$\alpha$ &-0.3 &0.0 &0.1 &0.2 &0.3 &0.4 &0.5 &0.6 &0.7 &0.8&exp.\\ \hline\hline 
 $\omega^{C}$ & 615 & 548 & 526 & 504 & 483 & 461 &
439 & 418 & 397 & 376& \\
 $\omega^{B}$ & 872 & 797 & 773 & 748 & 724 & 699 & 675 & 651 &
 626 & 602 &  \\ \hline
$\chi^{C}$ &0.165 &0.140 &0.132 &0.123 &0.114 &0.105 &0.096 &0.087
 &0.078 &0.069 & \\ 
$\chi^{B}$ &0.060 &0.053 &0.050 &0.046 &0.045 &0.042 &0.039 &0.037
 &0.034 &0.031 & \\ \hline
$\Delta M_{\Sigma_C}$ & 167 & 171 & 173 & 175 & 177 & 178 &
180 & 182 & 184 & 186 & 168\\
$\Delta M_{\Sigma_C}^\ast$ & 216 & 214 & 213 & 212 & 211
& 210 & 209 & 208 & 207 & 207 & 235\\ \hline
$\Delta M_{N}$ & -1187 & -1252 & -1274 & -1295 & -1316
& -1337 & -1358 & -1379 & -1399 & -1419 & -1344\\ \hline
$\Delta M_{\Lambda_B}$ & 3164 & 3172 & 3173 & 3176 & 3178
& 3181 & 3182 & 3185 & 3188 & 3191 & 3339$\pm$9 \\ \hline
\end{tabular}
\end{center}
\end{table}   

In the framework of collective coordinate quantization baryon
wavefunctions emerge as Wigner--$D$ functions of the collective
coordinates. These wavefunctions enter the computation of the
magnetic moments to be discussed in the next section. For baryons 
without heavy flavor components these wavefunctions read
\begin{equation}
{\bf \Psi}_N(I=J,I_{3},J,J_{3})={\cal N} D_{I_3,-J_3}^{(I=J)}(A)\,,
\label{LBwave}
\end{equation}
while for the baryons with a single heavy quark component we need 
to couple a diquark state ($J^{sol}$ is integer) of collective 
coordinates with unit occupation of the bound state that carries 
half--integer spin ($\vec{J}-\vec{J}^{sol}
=\vec{G}$) 
\begin{equation}
{\bf \Psi}_{B}(I,I_3,J,J_3) = {\cal N}^\prime\sum_{J_3^{sol},G_e}
C^{J\:J_3}_{I\:J_3^{sol},\:\frac{1}{2}\:G_{3}}
D_{I_3,-J_3^{sol}}^{I=J^{sol}}(A)|\frac{1}{2},G_3\rangle
\label{HBwave}
\end{equation}
to total spin according to eq.~(\ref{Grandspin}).
In the above equations, $C$ is a Clebsch--Gordan coefficient
while ${\cal N}$ and ${\cal N}^\prime$ are suitable
normalization constants.

\section{Electromagnetic current and results for magnetic moments}

In this section we compute the magnetic moments from matrix elements
of the electromagnetic current operator given in 
eqs.~(\ref{lightcurrent}) and~(\ref{Jh}). The defining equation
for the magnetic moment operator reads
\begin{equation}
\hat{\mu}=\frac{1}{2} \epsilon_{3ij}\int d^3r\, x_i
\left({\cal J}^{\rm light}_j+{\cal J}^{\rm heavy}_j\right)\,.
\label{defmu}
\end{equation}
We now substitute the field configurations~(\ref{Sol}), (\ref{PQi}),
(\ref{rhoI}) and (\ref{heavyR}) into eqs.~(\ref{lightcurrent}) 
and~(\ref{Jh}). We then obtain the operator as sum of terms that
are products of integrals over the profile functions and operators
that act in the space of the collective coordinates ($A$) and/or
as creation and annihilation operators ($\rho$) for the heavy
meson bound state,
\begin{equation}
\hat{\mu}=\mu_{S,0}\alpha^2\Omega_3-\mu_{V,0}D_{33}(A)
+\mu_{S,1}\rho^{\dagger}\frac{\tau_3}{2}\rho
-\mu_{V,1}D_{33}(A)\rho^{\dagger}\rho\,.
\label{mm}
\end{equation}
We omit the energy argument of the spinor $\rho$, it is understood to be 
the bound state energy $\epsilon$ The first two terms in eq.~(\ref{mm}) do not
contain the bound state wavefunction and are those already considered 
for the two flavor version of the model in ref.~\cite{Me89}. Note that for 
convenience we factorized the moment of inertia, $\alpha^2$ for the isoscalar 
piece to make contact with the notation in refs.~\cite{Ku89,Oh91}.
From eq.~(\ref{lightcurrent}) we find for this isoscalar piece,
\begin{eqnarray}
\mu_{S,0}&=&-\frac{8\pi M_{\rm N}}{9\alpha^2}\int{dr r^2}
\Bigg\{\frac{2m_V^2}{g^2}\phi
+\left[\frac{N_C}{4\pi^2}-\frac{1}{g}(\gamma_1+\frac{3}{2}\gamma_2)\right]
F^\prime\sin^2{F}\cr
&&\hspace{2cm}
+\frac{1}{2g}\gamma_2\left(\sin{F}(G^\prime-\xi^\prime_1)
+F^\prime[G(1-\xi_1)+\xi_2]\right)\cr
&&\hspace{2cm}
+\frac{1}{2g}(\gamma_2+2\gamma_3)F^\prime 
(\xi_1-1+\cos{F})(1+G-\cos{F})\cr
&&\hspace{2cm}
+2d_1\left[F^\prime (\xi_1+\xi_2)
-\frac{2}{r}\sin{F}(2+G-\xi_1-2\cos{F})\right]\Bigg\}\,,
\label{mus0}
\end{eqnarray}
while the explicit form of the isovector piece reads
\begin{eqnarray}
\mu_{V,0}&=&\frac{8\pi M_{\rm N}}{3}\int{dr r^2} 
\Bigg\{f_\pi^2 \sin^2{F}-\frac{m_V^2}{g^2}\cos{F}(1+G-\cos{F})\cr
&&\hspace{2cm}
+\frac{1}{g}(\gamma_1 +\frac{3}{2}\gamma_2)\omega F^\prime\sin^2{F}
-\frac{1}{2g}(\gamma_2+2\gamma_3)F^\prime\omega \cos{F}(1+G-\cos{F})\cr
&&\hspace{2cm}
+\frac{1}{2g}\gamma_2 \sin{F}[\omega^\prime(1+G-2\cos{F})
-G^\prime\omega]\cr
&&\hspace{2cm}
+2d_1\left[F^\prime\omega
+\frac{2}{r}\omega\sin{F}\cos{F}\right]\Bigg\}\,.
\label{muv0}
\end{eqnarray}
For the parameter set~(\ref{parameters}) we find numerically
$\mu_{S,0}=0.43\,{\rm n.m}$ and $\mu_{V,0}=6.70\,{\rm n.m.}$.
On the other hand $\mu_{S,1}$ and $\mu_{V,1}$ are bilinear in 
the heavy meson bound state wavefunctions.
Again, there are isoscalar
\begin{eqnarray}
\mu_{S,1}&=&-\frac{8M_N}{3}\left({\cal C}
-\frac{1}{6}\left(1-\alpha\right)\right)\int{dr r^2}
\Bigg\{\Phi^2\left(1+\frac{R_\alpha}{2}\right) \cr
&&+\frac{1}{2}\left[\Psi_0\left(R_{\alpha}\Psi_0
+\left(\epsilon-\frac{\alpha\omega}{2}\right)r\Psi_2\right)
+\Psi_2^2\left(1+\frac{R_\alpha}{2}\right)
-\Psi_1\left(\Psi_2^\prime r+\Psi_2+R_\alpha\Psi_1\right)\right]\cr
&&-\frac{d}{2}\Psi_0\left(\Psi_1\sin{F}-\Psi_2\frac{F^\prime r}{2}\right)
-\frac{\sqrt{2}c}{m_V g}
\Phi\left(\Psi_2\omega^\prime r-2\Psi_0G^\prime\right)\Bigg\}\,,
\label{mus1}
\end{eqnarray}
and isovector contributions 
\begin{eqnarray}
\mu_{V,1}&=&\frac{2M_N}{3}\left(1-\alpha\right)
\int{dr r^2 }\cos{F}
\Bigg\{\Phi_2^2\left(1+\frac{R_\alpha}{2}\right)
\cr
&&+\frac{1}{2}\left[-\Psi_0\left(R_\alpha\Psi_0
+\left(\epsilon-\frac{\alpha\omega}{2}\right)r\Psi_2\right)
+\Psi_2^2\left(1+\frac{R_\alpha}{2}\right)
+\Psi_1\left(\Psi_2^\prime r+\Psi_2+R_\alpha\Psi_{1}\right)\right]\cr
&&+d\,\Psi_0\Psi_1\sin{F}-\frac{\sqrt{2}c}{m_V g}
\Phi\left(\Psi_2\omega^\prime r-2\Psi_0G^\prime\right)\Bigg\}\cr
&&+\frac{dM_N}{3}\int{dr r^2}
\Big\{r\left(\Psi_0\Psi_2^\prime-\Psi_0^\prime\Psi_2\right)
+\Psi_0\Psi_2+2R_\alpha\Psi_0\Psi_1
\cr &&\hspace{3cm}
+2r\left(\epsilon-\frac{\alpha\omega}{2}\right)\Psi_1\Psi_2
-2Mr\sin{F}\Psi_2\Phi\Big\}\,,
\label{mus2}
\end{eqnarray}
with $R_{\alpha}=$ $\cos{F}-1+\alpha\left(1+G-\cos{F}\right)$. The factor
$M_N=940{\rm MeV}$ enters because we measure the magnetic moments
in nucleon magnetons [n.m.]. Numerical results
for both the charm and the bottom sectors are displayed in 
table~\ref{table_2}. 
\begin{table}[htb]
\caption{\label{table_2}
Numerical results for the coefficients in the magnetic moment 
operator~(\ref{mm}) that contain the bound state wavefunctions.
Data are given units of nucleon magnetons (n.m.). See also
table~\ref{table_1}.}
\begin{center}
\begin{tabular}{|r|c|c|c|c|c|c|c|c|c|c|c|} \hline
$\alpha$ &-0.3 &0.0 &0.1 &0.2 &0.3 &0.4 &0.5 &0.6 &0.7 &0.8\\ \hline
& \multicolumn{10}{|c|}{Charm Sector}\\  \hline
$\mu_{S,1}$ & -0.168 &-0.186 &-0.191 &-0.197 &-0.203 &-0.208 
&-0.214 &-0.219 &-0.225 &-0.230 \\
$\mu_{V,1}$ &-0.037 &-0.037 &-0.038 &-0.038 &-0.038 &-0.039 &-0.039
 &-0.040 &-0.040 &-0.041 \\ \hline 
& \multicolumn{10}{|c|}{Bottom Sector}\\  \hline
$\mu_{S,1}$ & 0.070 & 0.063 & 0.061 & 0.059 & 0.057 & 0.054 & 0.052 &
 0.050 & 0.048 & 0.046 \\
$\mu_{V,1}$ &-0.006 &-0.006 &-0.005&-0.004&-0.003 &-0.003&-0.002
 &-0.002 &-0.001 &-0.001 \\ \hline
\end{tabular}
\end{center}
\end{table}

We now sandwich the operator~(\ref{mm}) between the states
listed in eqs.~(\ref{LBwave})
and~(\ref{HBwave}) to obtain the magnetic moment of a given baryon
as linear combinations of the functionals $\mu_{S,0}$, ...,$\mu_{V,1}$.
This yields~\cite{Ku89,Oh91}:
\begin{eqnarray}
\mu(p) &=& \frac{1}{2}\mu_{S,0}+\frac{1}{3}\mu_{V,0} \nonumber\\
\mu(n) &=& \frac{1}{2}\mu_{S,0}-\frac{1}{3}\mu_{V,0} \nonumber\\
\mu(\Lambda_{Q}) &=& \frac{1}{2}\tilde{\mu}_{S,1} \nonumber\\
\mu(\Sigma_{Q}^{I_{3}= +1}) &=& \frac{2}{3}\mu_{S,0}
+\frac{1}{3}\mu_{V,0}-\frac{1}{6}\tilde{\mu}_{S,1}+\frac{1}{3}\mu_{V,1}\cr
\mu(\Sigma_{Q}^{I_{3}= 0 })&=&\frac{2}{3}\mu_{S,0}
-\frac{1}{6}\tilde{\mu}_{S,1}\nonumber\\
\mu(\Sigma_{Q}^{I_{3}= -1}) &=&\frac{2}{3}\mu_{S,0}
-\frac{1}{3}\mu_{V,0}-\frac{1}{6}\tilde{\mu}_{S,1}-\frac{1}{3}\mu_{V,1}\cr
\nonumber\\
\mu(\Sigma_{Q}^{I_{3}= 0}\rightarrow\Lambda_{Q}) &=&
-\frac{1}{3}\mu_{V,0}-\frac{1}{3}\mu_{V,1}\,.
\label{muHB}
\end{eqnarray}
The quantity $\tilde{\mu}_{S,1} = \chi\mu_{S,0}-\mu_{S,1}$ arises from the
quantization rule $\alpha^2\vec{\Omega}=\vec{J}^{sol}+\chi\vec{G}$. 
For the proton and the neutron only $\mu_{S,0}$ and $\mu_{V,0}$
enter. We immediately get $\mu(p)= 2.45\, {\rm n.m.} $ and 
$\mu(n)= -2.02\, {\rm n.m.}$ 
which agrees reasonably with the experimental data 
$\mu(p)= 2.79\, {\rm n.m.} ({\rm exp.})$ and 
$\mu(n)= -1.91\, {\rm n.m.}({\rm exp.})$.
As is well known~\cite{Me89}, the isoscalar combination 
$\mu_{S,0}=\mu(p)+\mu(n)$ is predicted somewhat too small in this
model, $0.43\, {\rm n.m.}$ vs. $0.88\, {\rm n.m.} ({\rm exp.})$
while the isovector combination
$(2/3)\mu_{V,0}=\mu(p)-\mu(n)$ is off by only a few percent,
$4.47\, {\rm n.m.}$ vs. $4.70\, {\rm n.m.} ({\rm exp.})$.

\begin{figure}[tb]
\centerline{
\epsfig{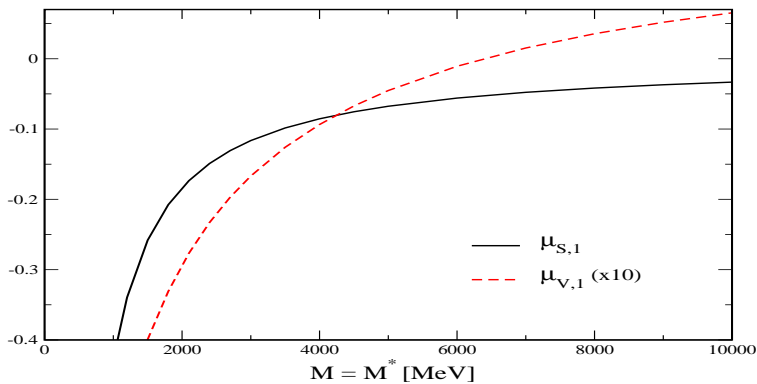}}
\caption{\label{fig1}Coefficients $\mu_{S,1}$ and $\mu_{V,1}$
in the magnetic moment operator~(\ref{mm}) that are bilinear
in the heavy meson wavefunction as functions of the heavy
meson masses for $\alpha=0$ in the charm sector, {\it i.e.}
for ${\cal C}=2/3$.}
\end{figure}
In figure~\ref{fig1} we show the numerical results for the 
coefficients $\mu_{S,1}$ and $\mu_{V,1}$
as functions of (equal) heavy meson masses. Obviously these coefficients
tend to zero as the mass increases\footnote{Note that the displayed
$\mu_{V,1}$ is multiplied by a factor 10 for clarity. For 
$M=M^*\gapeq6{\rm GeV}$ the value $\mu_{V,1}=0$ 
is consistent with the
numerical accuracy.}. This behavior is expected from the heavy mass limit 
since properties, such as the magnetic moments, that are related to spin 
of the heavy quark are suppressed by inverse powers of the heavy
quark mass. Note that also the hyperfine parameter $\chi$ tends to zero
in the heavy limit~\cite{Sch95a} and thus the reduction of 
$\mu_{S,1}$ implies that of $\tilde{\mu}_{S,1}$.
To nevertheless obtain a non--vanishing result in the heavy limit 
the authors of ref.~\cite{Oh95} added an extra term 
with an undetermined parameter to ${\cal L}_{\rm heavy}$
that is linear in the photon field and thus does only contribute
to electromagnetic properties of the heavy baryons~\cite{Cheng93}. It 
was argued in ref.~\cite{Oh95} that this term would be necessary to 
describe the radiative decays $Q\to P\gamma$. This term introduces 
an additional, so far undetermined parameter which may be interpreted 
as the intrinsic magnetic moment of the heavy meson field. Adopting 
a canonical value, $1/2M$, leads to 
$\mu_{\Lambda_c}\approx 0.37{\rm n.m.}$~\cite{Sa94}, a value 
considerably larger than our result. It is important to note that the 
model we consider is capable of describing such radiative decays~\cite{Ja95} 
because it is formulated for finite heavy meson masses rather than 
in the strict heavy limit. Thus there is no need for 
any additional photon coupling in our model. The
decrease of the $\mu_{S,1}$ and $\mu_{V,1}$ with the increase of 
the heavy meson mass can also be observed from table~\ref{table_2}
where we list these quantities for the physical cases as functions
of the undetermined parameter $\alpha$. In this case, of course, 
the pseudoscalar and vector meson masses are different, {\it cf.}
eq.~(\ref{hPar1}).

We see that the coefficients $\mu_{S,1}$
and $\mu_{V,1}$ that are bilinear in the heavy meson wavefunctions
exhibit quite a pronounced dependence on the undetermined 
parameter $\alpha$. However, this dependence does not propagate
to the magnetic moments of the heavy baryons as seen in 
tables~\ref{table_3} and~\ref{table_4}. One reason is that
the total magnetic moments are dominated by $\mu_{S,0}$
and $\mu_{V,0}$, the coefficients that contain only the light 
mesons profile functions. The other reason is that the
combination $\tilde{\mu}_{S,1}=\chi\mu_{S,0}-\mu_{S,1}$
essentially stays constant due to the decrease of $\chi$ 
with $\alpha$, {\it cf.} table~\ref{table_1}.

\begin{table}[htb]
\caption{\label{table_3}
Magnetic moments of heavy baryons in the charm sector given
in units of nuclear magnetons [n.m.]}
\begin{center}
\begin{tabular}{|c|c|c|c|c|c|}\hline
$\alpha$ &${\quad\Lambda_{C}\quad}$ & ${\quad\Sigma_{C}^{++}\quad}$ & 
${\quad\Sigma_{C}^{+}\quad}$ & ${\quad\Sigma_{C}^{0}\quad}$
& ${\Sigma_{C}^{+}\rightarrow \Lambda_{C}}$\\ \hline\hline
-0.3 & 0.12 & 2.46 & 0.25 & -1.96 & -2.26 \\ \hline
0 & 0.12 & 2.45 & 0.25 & -1.96 & -2.26 \\ \hline
0.3 & 0.13 & 2.45 & 0.24 & -1.96 & -2.26 \\ \hline
0.6 & 0.13 & 2.45 & 0.24 & -1.96 & -2.26 \\ \hline
\end{tabular}
\end{center}
\end{table}
 
\begin{table}[htb]
\caption{\label{table_4}
Magnetic moments of
heavy baryons in the bottom sector given in units of nuclear magnetons [n.m.]}
\begin{center}
\begin{tabular}{|c|c|c|c|c|c|}\hline
$\alpha$ &${\quad\Lambda_{B}\quad}$ & ${\quad\Sigma_{B}^{+}\quad}$ & 
${\quad\Sigma_{B}^{0}\quad}$ & ${\quad\Sigma_{B}^{-}\quad}$
& ${\Sigma_{B}^{0}\rightarrow \Lambda_{B}}$\\ \hline\hline
-0.3 &-0.02 & 2.52 & 0.29 & -1.93 & -2.24 \\ \hline
0 &-0.02 & 2.52 & 0.29 & -1.93 & -2.24 \\ \hline
0.3 &-0.02 & 2.52 & 0.29 & -1.94 & -2.24 \\ \hline
0.6 &-0.02 & 2.52 & 0.29 & -1.94 & -2.23 \\ \hline
\end{tabular}
\end{center}
\end{table}

In table~\ref{table_5} we show the results for the magnetic 
moments obtained in bound state approach to the Skyrme model.
For the charm sector these results are the same as those 
denoted\footnote{Note that the authors of ref.~\cite{Oh91}
normalize their computed magnetic moments to that of the
nucleon.} ``SET II'' in table~7 of ref.~\cite{Oh91}. The
calculation in the bottom sector is that of the kaon sector
with the replacement $M_{\rm K}\to5279{\rm MeV}$.
\begin{table}[htb]
\caption{\label{table_5}
Magnetic moments (in nuclear magnetons) of heavy baryons using 
the bound state approach to the Skyrme model~\cite{Oh91}.
In this model we find $\mu_p=1.97{\rm n.m.}$ and $\mu_n=-1.24{\rm n.m.}$.}
\begin{center}
\begin{tabular}{|c|c|c|c|c|c|}\hline
&${\quad\Lambda_{Q}\quad}$ & ${\quad\Sigma_{Q}^{I_3=1}\quad}$ &
${\quad\Sigma_{Q}^{I_3=0}\quad}$ & ${\quad\Sigma_{Q}^{I_3=-1}\quad}$
& ${\Sigma_{Q}^{I_3=0}\rightarrow \Lambda_{Q}}$\\ \hline\hline
$Q=C$ &0.20 & 1.96 & 0.42 & -1.12 & -1.67 \\ \hline
$Q=B$ &-0.21 & 2.22 & 0.56 & -1.11 & -1.54 \\ \hline
\end{tabular}
\end{center}
\end{table}
Besides the lower scale for the magnetic moments of the light
baryons, the essential difference is that the contribution
of the heavy meson bound states to the magnetic moments does
not decrease as the heavy meson mass increases. Of course, 
that is expected, as the model of ref.~\cite{Oh91} does not
reflect the heavy quark spin symmetry.

We finally would like to compare our predictions with 
those of other descriptions for heavy baryons. Analyses of
QCD spectral sum rules yield~\cite{Zhu97}:
$\mu_{\Lambda_c}=(0.15\pm0.05){\rm n.m.}$,
$\mu_{\Sigma_c^{++}}=(2.1\pm0.3){\rm n.m.}$,
$\mu_{\Sigma_c^{+}}=(0.23\pm0.03){\rm n.m.}$, and
$\mu_{\Sigma_c^{0}}=(-1.6\pm0.2){\rm n.m.}$. 
These values agree nicely with our results. On the other hand, this
result for the magnetic moment of $\Lambda_c$ is 
smaller than the one obtained from assuming a canonical 
intrinsic magnetic moment of the heavy meson fields~\cite{Sa94}.
In our approach those four magnetic moments ($\mu_{\Lambda_c}$,
$\mu_{\Sigma_c^{++}}$, $\mu_{\Sigma_c^{+}}$ and $\mu_{\Sigma_c^{0}}$)
contain the isoscalar piece $\mu_{{\Sigma}_Q^{I_3=0}}$ which is 
dominated by the light meson isoscalar contribution, $\mu_{S,0}$.
We therefore compare our results for the transition magnetic moments, 
$\mu_{{\Sigma}_Q^{I_3=0}\to\Lambda_Q}$ which do not contain
$\mu_{{\Sigma}_Q^{I_3=0}}$, to results from light cone QCD sum
rules~\cite{Al01}: $\mu_{{\Sigma}_C\to\Lambda_C}=-(1.5\pm0.4){\rm n.m.}$
and $\mu_{{\Sigma}_B\to\Lambda_B}=-(1.6\pm0.4){\rm n.m.}$. We find that our
predictions are only slightly larger (in magnitude), in addition, those sum 
rule results have sizable error bars.

\section{Summary}

In this study we have employed the bound state approach to compute magnetic 
moments of heavy baryons with a single heavy quark. In this approach the heavy
baryon is constructed from a heavy meson field that is bound in the background 
field of a light baryon. The latter emerges as a soliton of light meson fields.
In the model, that we consider here, this soliton contains both light 
pseudoscalar and light vector meson fields. This extension of the original 
Skyrme model is known to reasonably describe the phenomenology of light baryons. 
This is particularly the case for the magnetic moments of proton and neutron. 
For the heavy sector we have adopted a relativistic Lagrangian which does not 
directly reflect the heavy quark symmetry. Rather this model embodies 
the physical values of the heavy meson masses. However, in the limit of 
infinitely large heavy meson masses it properly reflects the heavy spin flavor
symmetry. We have then introduced and canonically quantized the collective 
coordinates that parameterize the spin--flavor orientation of the soliton 
and the bound state wavefunction to generate baryon states. 

We have extracted the operator of the electromagnetic current from
the electromagnetically gauged action. To compute baryon magnetic
moments we have sandwiched the appropriate combination of this operator
between baryon states. For the magnetic moments of the heavy baryons 
experimental data do not yet exist, thus we have compared our 
results with to predictions of other models. Specifically, results
are available for spectral sum rule and QCD light cone sum rule 
analyses. Although our prediction for the transition magnetic 
moments, $\mu_{{\Sigma}_Q^{I_3=0}\to\Lambda_Q}$ is slightly larger 
than in the QCD sum rule approach, the overall picture is that the predictions
in these two descriptions for heavy baryon properties 
are very similar. On the other hand, we have seen
that the bound state approach to the $SU_F(N)$ Skyrme model, which does 
not respect the heavy spin flavor symmetry, gives significantly different
predictions to the magnetic moments of heavy baryons with a single quark.
Not surprisingly, these differences are quite distinct for the magnetic moment 
of $\Lambda_Q$ which is most sensitive to the heavy quark contribution
to the magnetic moments.

\section*{Acknowledgements}
We are grateful to J. Schechter for fruitful discussions.
\appendix
\section*{Appendix}

In this appendix we list the explicit expressions for the functionals 
of various meson profiles that parameterize the collective coordinate
Lagrangian~(\ref{Lcoll}). These functionals have already been
reported in the literature~\cite{Me89,We96,Sch95a}. However,
because notations do vary within those papers, we list them here 
for completeness.

First we focus on quantities that only involve the light
pseudoscalar and light vector meson profiles, {\it cf.}
eq.~(\ref{Sol}).  The classical mass is given by
\begin{eqnarray}
M_{\rm cl}&=&4\pi\int dr\left\{\frac{f_\pi^2}{2}(F^{\prime2}r^2+2\sin{F})
+m_\pi^2f_\pi^2\left(1-\cos{F}\right)\right.\cr 
&&\hspace{1.5cm}
-\frac{r^2}{2g^2}(\omega^{\prime2}+m_V^2\omega^2)
+\frac{1}{g^2}[G^{\prime2}+\frac{G^2}{2r^2}(G+2)^2]\cr
&&\hspace{1.5cm}
+\frac{m_V^2}{g^2}(1+G-\cos{F})^2
+\frac{\gamma_1}{g}F^\prime\omega\sin^2{F}
-\frac{2\gamma_2}{g}G^\prime\omega\sin{F}\cr
&&\hspace{1.5cm}\left.
+\frac{\gamma_3}{g}F^\prime\omega G(G+2)
+\frac{1}{g}\left(\gamma_2+\gamma_3\right)F^\prime\omega
\left[1-2(G+1)\cos{F}+\cos^2{F}\right]\right\}\,.
\label{mcl}
\end{eqnarray}
Its variation gives the soliton profiles $F$, $G$ and~$\omega$. The
moment of inertia reads
\begin{eqnarray}
\alpha^2&=&
\frac{8\pi}{3}\int dr \Big\{f_\pi^2 r^2\sin^2{F}
-\frac{4}{g^2}(\varphi^{\prime2}+2\frac{\varphi^2}{r^2}+m_V^2\varphi^2)
+\frac{m_V^2}{2g^2}r^2[(\xi_1+\xi_2)^2+2(\xi_1-1+\cos{F})^2]\cr
&&\hspace{1.5cm}
+\frac{1}{2g^2}[(3\xi_1^{\prime2} +2\xi_1^\prime\xi_2^\prime+
\xi_2^{\prime2})r^2+2(G^2+2G+2)\xi_2^2+4G^2(\xi_1^2+\xi_1\xi_2
-2\xi_1-\xi_2+1)]\cr
&&\hspace{1.5cm}
+\frac{4}{g}\gamma_1\varphi F^\prime\sin^2{F}
+\frac{4}{g}\gamma_3\varphi F^\prime[(G-\xi_1)(1-\cos{F})
+(1-\cos{F})^2-G\xi_1]\cr
&&\hspace{1.5cm}
+\frac{2\gamma_2}{g}\left[\varphi^\prime\sin{F}(G-\xi_1+2-2\cos{F})
+\varphi \sin{F}(\xi_1^\prime-G^\prime)\right.\cr
&&\hspace{1.5cm}
\left.
+\varphi F^\prime\left(2+2\sin^2{F}+(\xi_1-G-2)\cos{F}
-2(\xi_1+\xi_2)\right)\right]\Big\}\,.
\label{mominer}
\end{eqnarray}
Again, by variation the profile functions $\xi_1$, $\xi_2$ and~$\varphi$
are obtained. The classical profile functions serve as source fields.

The quantities $I_\epsilon$ and $\chi$ are additionally functionals
of the heavy meson fields. For the bound state in the 
P--wave channel one obtains upon substitution of the
{\it ansatz} (\ref{PQi}) 
\begin{eqnarray}
I_\epsilon&=&\int dr r^2 \Bigg(\Phi^{\prime2}+
\left[M^2-\left(\epsilon-\frac{\alpha}{2}\omega\right)^2
+\frac{2}{r^2}\left(1+\frac{1}{2}R_\alpha\right)^2\right]\Phi^2
+M^{*2}\left[\Psi_1^2+\frac{1}{2}\Psi_2^2-\Psi_0^2\right]
\nonumber \\ &&
+\frac{1}{2}\left[\Psi_2^\prime-\frac{1}{r}\Psi_2\right]^2
+\frac{1}{r}R_\alpha\Psi_1\Psi_2^\prime
+\frac{1}{r^2}R_\alpha\left(\Psi_1+\Psi_2\right)\Psi_2
+\frac{1}{2r^2}R_\alpha^2\left(\Psi_1^2+\frac{1}{2}\Psi_2^2\right)
\nonumber \\ &&
-\left[\Psi_0^\prime-
\left(\epsilon-\frac{\alpha}{2}\omega\right)\Psi_1\right]^2
-\frac{1}{2}\left[\frac{R_\alpha}{r}\Psi_0
+\left(\epsilon-\frac{\alpha}{2}\omega\right)\Psi_2\right]^2
\nonumber \\ &&
-d\Bigg\{\frac{2}{r}{\rm sin}F\left[\Psi_2\Psi_0^\prime
-\frac{R_\alpha}{r}\Psi_0\Psi_1
-\left(\epsilon-\frac{\alpha}{2}\omega\right)\Psi_1\Psi_2\right]
\nonumber \\ && \hspace{1cm}
+\frac{F^\prime}{r}\left[\frac{r}{2}
\left(\epsilon-\frac{\alpha}{2}\omega\right)\Psi_2^2
-\left(1-{\rm cos}F\right)\Psi_0\Psi_2\right]\Bigg\}
+2Md\left[F^\prime\Psi_1-\frac{{\rm sin}F}{r}\Psi_2\right]\Phi
\nonumber \\ &&
+\frac{2\sqrt{2}cM}{gm_V}\left[2\omega^\prime\Psi_0\Psi_1
-\frac{2G^\prime}{r}\Psi_1\Psi_2
+\frac{G}{2r^2}\left(G+2\right)\Psi_2^2\right]
\nonumber \\ &&
-\frac{4\sqrt{2}c}{gm_V}\Bigg\{
\frac{1}{r^2}\left(\epsilon-\frac{\alpha}{2}\omega\right)
\left[G\left(G+2\right)\Psi_1-rG^\prime\Psi_2\right]\Phi
-\frac{\omega^\prime}{r}\left[1+\frac{R_\alpha}{2}\right]\Psi_2
\nonumber \\ && \hspace{2cm}
+\frac{1}{r^2}\left[G\left(G+2\right)\Phi^\prime
+G^\prime\left(2+R_\alpha\right)\Phi\right]\Psi_0\Bigg\}
\Bigg)\,.
\label{pwlag}
\end{eqnarray}
Here a prime indicates a derivative with respect to the radial
coordinate $r$. Furthermore the abbreviation
$R_\alpha={\rm cos}F-1+\alpha\left(1+G-{\rm cos}F\right)$
has again been used. The functional $I_\epsilon$ leads to the
equations of motion for the profile functions $\Phi$, $\Psi_0$,
$\Psi_1$, and $\Psi_2$ of the fluctuating heavy meson fields.
In these equations the classical fields generate the binding 
potential. The solution to these equations provides the bound state
energy, $\epsilon$ and the bound state wavefunctions that
are subsequently normalized according to eq.~(\ref{PNorm}).

Finally, we present the explicit expressions for the
hyperfine splitting parameter for the bound state in the
$P$--wave, {\it cf.} section 4. 
For convenience we employ additional abbreviations with 
regard to the light meson profiles defined in 
eqs (\ref{Sol}) and (\ref{rhoI})
\begin{eqnarray}
V_1&=&{\rm cos}F-\alpha\left(\xi_1-1+{\rm cos}F\right)\ ,
\nonumber \\
V_2&=&1-\alpha\left(\xi_1+\xi_2\right) \ .
\nonumber
\end{eqnarray}
The explicit expression for the P--wave hyperfine parameter,
which enters the mass formula for the even parity heavy 
baryon (\ref{Hmass}), reads
\begin{eqnarray}
\chi&=&\frac{2}{3}\int_0^\infty dr\ r^2\ \rho_{\chi}(r)
\label{eq3} \\
\rho_{\chi}(r)\hspace{-5pt}&=& \hspace{-4pt}
\left[\left(\epsilon-\frac{\alpha}{2}\omega\right)\left(V_2-2V_1\right)
-\frac{2\alpha}{r^2}\left(2+R_\alpha\right)\varphi\right]\Phi^2
\nonumber \\ && \hspace{-4pt}
+\left(2V_1+V_2\right)\left[\left(\epsilon-\frac{\alpha}{2}\omega\right)
\Psi_1-\Psi_0^\prime\right]\Psi_1
\nonumber \\ && \hspace{-4pt}
-\frac{1}{2}\left(V_2\Psi_2+\frac{4\alpha}{r}\varphi\Psi_0\right)
\left[\left(\epsilon-\frac{\alpha}{2}\omega\right)
\Psi_2+\frac{R_\alpha}{r}\Psi_0\right]
\nonumber \\ && \hspace{-4pt}
+\frac{2\alpha}{r}\varphi\Psi_1
\left(\Psi_2^\prime+\frac{1}{r}\Psi_2+\frac{R_\alpha}{r}\Psi_1\right)
-\frac{\alpha}{r^2}\left(2+R_\alpha\right)\varphi\Psi_2^2
+4Md\ {\rm sin}F\Phi\Psi_0
\nonumber \\ && \hspace{-4pt}
-\frac{d}{r}
\left\{{\rm sin}F\left[\left(2+R_\alpha+V_1\right)\Psi_1\Psi_2
-\frac{4\alpha}{r}\varphi\Psi_0\Psi_1\right]
+F^\prime\left[\frac{r}{4}V_2\Psi_2^2+
2\alpha\varphi\Psi_0\Psi_2\right]\right\}
\nonumber \\ && \hspace{-4pt}
-\frac{4\sqrt2cM}{gm_V}\left\{
\left(3\xi_1^\prime+\xi_2^\prime\right)\Psi_0\Psi_1
+\frac{G}{r}\left(2-2\xi_1-\xi_2\right)\Psi_0\Psi_2
+\frac{2}{r}\varphi^\prime\Psi_1\Psi_2
+\frac{1}{r^2}\varphi\Psi_2^2\right\}
\nonumber \\ && \hspace{-4pt}
-\frac{4\sqrt2c}{gm_V}
\Bigg\{\left(\epsilon-\frac{\alpha}{2}\omega\right)
\left(\frac{4}{r^2}\varphi\Psi_1+
\frac{2}{r}\varphi^\prime\Psi_2\right)\Phi
+\left(V_1-\frac{V_2}{2}\right)
\left[\frac{G}{r^2}\left(G+2\right)\Psi_1
-\frac{G^\prime}{r}\Psi_2\right]\Phi
\nonumber \\ && \hspace{-4pt}
+\frac{G}{r^2}\left(2+R_\alpha\right)
\left(2\xi_1+\xi_2-2\right)\Phi\Psi_1
+\frac{2}{r^2}\left[2\varphi\Phi^\prime
+\left(2+R_\alpha\right)\varphi^\prime\Phi
-\alpha G^\prime\varphi\Phi\right]\Psi_0
\nonumber \\ && \hspace{-4pt}
-\frac{1}{r}\left[\left(G+2\right)\xi_2\Phi^\prime
+\left(1+\frac{1}{2}R_\alpha\right)
\left(\xi_1^\prime+\xi_2^\prime\right)\Phi
-\alpha\omega^\prime\varphi\Phi\right]\Psi_2\Bigg\}\,.
\end{eqnarray}
Substituting the bound state profiles as well as the soliton
yields numerical results which are used to compute the
heavy baryon spectrum according to eq.~(\ref{Hmass}).

\end{document}